\documentclass[prl,twocolumn,preprintnumbers,amsmath,amssymb,bibliography]{revtex4-1}

\usepackage{graphicx}
\usepackage{color}

\begin{document}

\title{Phonon mediated interlayer conductance in twisted graphene bilayers}
\author{V. Perebeinos, J. Tersoff, and Ph.\,\,Avouris}
\affiliation{IBM T.J. Watson Research Center, Yorktown Heights, NY 10598, USA}
\date{\today}

\begin{abstract}

Conduction between graphene layers is suppressed by momentum conservation whenever the layer stacking has a rotation.  Here we show that phonon scattering plays a crucial role in facilitating interlayer conduction.  The resulting dependence on orientation is radically different than previously expected, and far more favorable for device applications. At low temperatures, we predict diode-like current-voltage characteristics due to a phonon bottleneck. Simple scaling relationships give a good description of the conductance as a function of temperature, doping, rotation angle, and bias voltage, reflecting the dominant role of the interlayer beating phonon mode.
\end{abstract}

\maketitle

Graphene has generated broad excitement both for fundamental science and for its potential applications in technology \cite{Geim07}.  Attention has turned increasingly to bilayer and few layer graphene \cite{CastroNeto09}, because of their scientific richness and their promise in technological applications involving band-gap opening by an external electric field, high current-carrying capacity, and electro-optical coupling \cite{McCann,Ohta1,Castro,Oostinga,WangGap,TonyGap,KuzmenkoGap,XiaGap,WendyGap,Ferrari2}. Interlayer conductance is crucial in almost all such applications. However, most fabrication methods lead to ``twisted'' layer stacking, with a random angle of rotation between layers.  As a result, momentum conservation in certain respects  ``decouples'' the layers \cite{deHeer08,Faugeras,Sprinkle,Sprinkle2,Shallcross1,Mele1,McDonald1}.
Strictly speaking, the layers do couple, as evidenced by velocity renormalization
\cite{Santos,Trambly,Morell,Andrei1,EPAPS}. But the states near the Dirac point in one layer largely decouple from those near the rotated Dirac point in the other layer \cite{Shallcross1,Shallcross2,Mele1,McDonald1}, suppressing interlayer current \cite{McDonald1}.

Understanding the interlayer conduction has proven surprisingly subtle and difficult. Calculations of the interlayer conductance to date have required some phenomenological lifetime broadening, with the conductance depending on this broadening \cite{McDonald1}.  In the limit of small carrier scattering the incoherent transport picture will fail, and it becomes problematic to even define an interlayer conductance independent of the external contacts \cite{McDonald1}.  With a realistic scattering lifetime the conductance is well defined, but it's calculated value is  extraordinarily sensitive to even small details of the matrix-element modeling \cite{McDonald1,Shallcross2}.
Finite conductance is predicted even for twisted stacking \cite{McDonald1} due to Umklapp electronic coupling \cite{Mele1}, but the coupling decays exponentially with the size of the rotational supercell \cite{Shallcross1,Shallcross2,Mele1,McDonald1}. (Here ``supercell'' denotes the primitive cell of the Bravais lattice of the commensurate rotated bilayer.) This would pose a serious obstacle to many device applications, where some reliable lower bound on the interlayer conductivity is essential. Experimentally, some interlayer transport is observed \cite{Yuming_APL,Kuroda_ACS}, but the mechanism remains unclear.  Extrinsic scattering mechanisms could also relax momentum conservation, especially for defect-rich substrates.

Here we show that essentially all of the conceptual and computational problems of twisted bilayers are resolved by including phonon scattering explicitly in the calculation of interlayer transport. Then incoherent transport is assured, no phenomenological broadening is required, and the calculated conductance is not so sensitive to the matrix-element model.  The phonon-mediated conductance arises primarily from the vibrational ``beating'' mode of the bilayer, in which out-of-plane acoustic (flexural) modes of the two layers are 180$^{\circ}$ out of phase, so the layers vibrate against each other.   We introduce a simple  tight-binding Hamiltonian that can describe the electron scattering by this mode for arbitrary rotation.

The resulting dependence on rotation angle is quite different than previously expected, and more favorable for device applications.  In particular, the extraordinary sensitivity to rotation angle disappears, replaced by a smooth and mild dependence;  and the interlayer conductance is never very small at room temperature. Thus device behavior can be robust despite random rotation angles. Simple scaling relationships give a good description of the conductance as a function of temperature, doping, rotation angle, and bias.  Also, at low temperature we find a diode-like turn-on of the conductance with increasing voltage.

The phonon mediated current between the layers is given by \cite{Datta}:
\begin{eqnarray}
 I_{ph}= \frac{4\pi e}{\hbar }\sum
M^{ks}_{k^{\prime}s^{\prime}\mu} \left[f(E_{ks})-f(E_{k^{\prime}s^{\prime}}+eV)\right]
\label{eqs1}
\end{eqnarray}
where the sum is over $k,k^{\prime},s,s^{\prime},\mu$ and spin degeneracy has been included. Here $k$ and $s$ label 2D electron wavevector and band, respectively, $\mu$ labels phonon bands, $f$ is the Fermi function, and $V$ is the voltage bias between the layers.
To first order,
\begin{eqnarray}
M^{ks}_{k^{\prime}s^{\prime}\mu} =&&\vert \langle \Psi_{ks}|H_{e-ph}^{\mu}|\Psi_{k^{\prime}s^{\prime}} \rangle \vert^2
[n_{q\mu}\delta(E_{k^{\prime}s^{\prime}}-E_{ks}+\hbar\omega_{q\mu})
\nonumber\\
&&+(1+n_{-q\mu})\delta(E_{k^{\prime}s^{\prime}}-E_{ks}-\hbar\omega_{-q\mu})]
\label{eqs2}
\end{eqnarray}
where $n$ is the Bose-Einstein occupancy, and $\langle\Psi_{ks}|H_{e-ph}^{\mu}|\Psi_{k^{\prime}s^{\prime}}\rangle$ is the electron-phonon matrix element for carrier scattering from state $\Psi_{ks}$ to $\Psi_{k^{\prime}s^{\prime}}$ by emitting or absorbing a phonon $\hbar\omega_{q\mu}$ in branch $\mu$ with wavevector  $q=k-k^{\prime}$. The electron energy $E_{k}$ in each monolayer is described by the tight binding model with the in-plane first neighbor hopping \cite{CastroNeto09} $t=3.1$ eV; and only out-of-plane phonons are included in evaluating $H_{e-ph}$. The conductance is $G=I/(VA)$, where $A$ is the area.

\begin{figure}[hb]
\centering
\scalebox{1.0}[1.0]{
\includegraphics[width=3.5in]{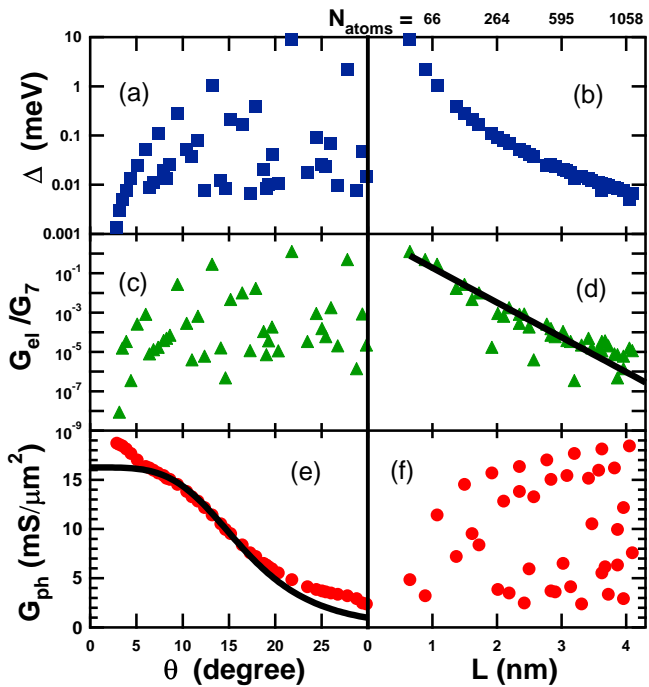}}
\caption{(Color  online)  Dependence of bilayer coupling and conductance on rotation.  Calculated band splitting $\Delta$ is shown vs (a) rotation angle $\theta$, and (b) the corresponding period $L$ of the rotational supercell.  Electronic conductance (i.e. calculated omitting phonon scattering) at $T=0$, with doping level $E_F=0.26$ eV in both layers \protect\cite{ftnote2},
is shown vs (c) angle $\theta$ and (d) period $L$. See text for reference conductance $G_7$ and for the solid line in (d).  Phonon mediated conductance at room temperature and same doping is shown vs (e) $\theta$ and (f) $L$. The solid black curve in (e) is given by Eq.~(\protect{\ref{eqscale}}) with only one adjustable parameter, $g$=0.34  eV/\AA.  Deviations at angles $\lesssim 2k_F/K\simeq 3^{\circ}$ are expected, see text.}
\label{figD}
\end{figure}

To render the calculation computationally tractable requires a simple tight-binding model such as that of Ref.~\cite{Ando1}.   That model or any reasonable model would suffice for demonstrating the qualitative effects of electron-phonon coupling, which can be directly understood from the requirements of momentum and energy conservation in the electron+phonon system.  However, to obtain reasonable quantitative accuracy, we desire a model that correctly describes the dependence on both rotation and interlayer spacing, insofar as these are known from LDA calculations.  Several authors have examined this problem under the constraint of fixed interlayer spacing \cite{Shallcross2,Trambly}, which is sufficient in the absence of phonons.  The ``band splitting'' $\Delta$ in the bilayer \cite{Shallcross1} is often used to conveniently represent the degree of interlayer Dirac-point coupling, and hence the interlayer conductance \cite{Mele1,McDonald1}.  ($\Delta$ is the energy from the second valence band at K to the second conduction band, shown in Fig.~3 of Ref.~\cite{Mele1}.) We require that the model also reproduces the few available LDA results \cite{EPAPS} for the dependence of $\Delta$ on layer spacing.  This rules out the existing models, with the possible exception of that in Ref.~\cite{Shallcross2}, which is too cumbersome for our purposes.
We therefore adapt the simplest available model \cite{Ando1} with a phenomenological lateral screening length $\lambda_{xy}$,
\begin{eqnarray}
&&t_{ij}=t_{\perp}\exp{\left(-\frac{r_{ij}-h_0}{\lambda_z}\right)}\exp{\left(-\left(\frac{\xi_{ij}}{\lambda_{xy}}\right)^{\alpha}\right)}
\label{eqtij}
\end{eqnarray}
where $h_0=3.35$ \AA \ is the equilibrium interlayer distance, $r_{ij}$ is distance between carbon atoms in adjacent layers, and $\xi_{ij}=((x_i-x_j)^2+(y_i-y_j)^2)^{1/2}$ is the in-plane distance between the atoms. Here we take $t_{\perp}=0.4$ eV.  We obtain a reasonable fit to the available LDA results for $\Delta$ in rotated bilayers \cite{Shallcross1} and dependence on interlayer spacing \cite{EPAPS} using $\lambda_z=0.6$ \AA, \ $\lambda_{xy}=1.7$ \AA \ and $\alpha$=1.65. The model also reasonably well reproduces LDA results for the Fermi velocity renormalization \cite{EPAPS}.


The phonon eigenvalues and eigenvectors are calculated using an atomistic valence force model \cite{PerebeinosTersoff} plus a Lennard-Jones (LJ) 6-12  potential \cite{Lad,LJparam} between atoms in adjacent layers. The electron-phonon Hamiltonian is obtained by expanding Eq.~(\ref{eqtij}) to linear order in the phonon displacements.

\begin{figure}[hb]
\centering
\scalebox{1.0}[1.0]{
\includegraphics[width=3.4in]{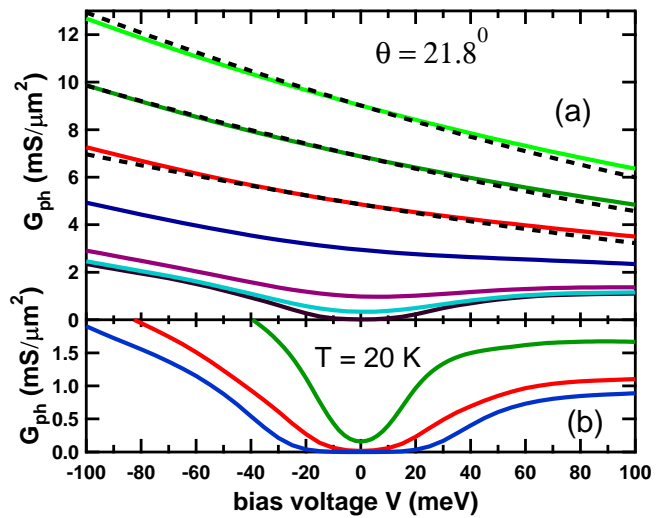}}
\caption{ (Color  online)   Bias dependence of the phonon-mediated conductance $G_{ph}$. (a) $G_{ph}$ for different temperatures $T$, $E_F$=0.26 eV and  $\theta=21.8^{\circ}$. From top to bottom, $T=$ 500 K (light green), 400 K (green), 300 K (red), 200 K (blue), 100 K (magenta), 60 K (cyan), 20K (black). The black dashed curves are single parameter fits described in text. This scaling breaks down at temperatures below the energy of the beating phonon mode. (b) $G_{ph}$ at $T=20$ K and $E_F$=0.26 eV for three different structures: from top to bottom, $\theta=13.2^{\circ}$ (green curve), $\theta=21.8^{\circ}$ (red curve), and $\theta= 27.8^{\circ}$ (blue curve).  The width of the plateau reflects the beating phonon energy (Fig.~\ref{figE}b).}
\label{figG}
\end{figure}

The results for the phonon mediated conductance are summarized in Fig.~\ref{figD}, and compared with the electronic conductances and band splittings.
We include all the commensurate structures up to $N_{atoms}=4L^2/a^2\approx1100$ atoms per rotational supercell.  The splitting  $\Delta$ is obtained from the low-energy effective 4$\times$4 Hamiltonian \cite{Mele1,McDonald1} calculated using Eq.~(\ref{eqtij}).  The specific commensurate structure is characterized by a graphene lattice vector \cite{Santos,Trambly,Mele1,Shallcross1}, so \textit{a priori} it need not follow a smooth curve when plotted against any single scalar variable. In Fig.~\ref{figD}(a) and (b) the splitting appears chaotic when plotted against angle.  However, a simple smooth trend emerges when plotted against the size of the supercell, reflecting the role of umklapp processes \cite{McDonald1,Mele1,Shallcross2,Shallcross1}).  As expected,  in Fig~\ref{figD}(c) and (d) we see  generally similar behavior for the calculated electronic conductance $G_{el}$ \cite{Geldetail}. Both $\Delta$ and $G_{el}$ decay with increasing supercell size, such that $G_{el}$ decreases by 7 orders of magnitude when  $L$ increases from 2 nm to 6 nm, which is typical of the range seen in bilayer graphene Moir\'e patterns \cite{Rong,Pong,Strocio,Hiebel,Varchon1,Andrei1}. If phonon scattering is omitted, the calculation of electronic conductance requires some phenomenological broadening, and is proportional to the value of this broadening \cite{McDonald1}.  We therefore give the electronic conductance relative to that of the smallest nontrivial rotated structure, i.e. $\sqrt{7}\times\sqrt{7}$, since the ratio is independent of the broadening. $G_7$ was estimated \cite{McDonald1} to be $G_7\approx 30$ mS/$\mu$m$^2$ for $E_F=0.26$ eV.

When phonon scattering is included, the behavior is radically different in two distinct ways.  First, there is a smooth dependence on rotation angle and {\it not} on supercell size, the opposite of the behavior without phonons.  Second, the maximum variation with angle (excluding ideal AB stacking and very small angles) is only about a factor of $10$, rather than the exponential variation seen without phonons in Fig.~\ref{figD}d.  Thus phonon-mediated conduction is expected to dominate except perhaps near orientations where the supercell is particularly small.

The exponential decay of electronic coupling $\Delta$ with supercell size $L$, shown in Fig.~\ref{figD}b, can be understood from previous work \cite{McDonald1,Mele1,Shallcross2,Shallcross1}. As discussed there, states near the K point of one layer are coupled to the rotated K point of the other layer by Umklapp process.  The larger the commensuration period $L$, the larger the reciprocal lattice vector $Q$ needed for Umklapp coupling. The coupling is related to the $Q$ Fourier component of the finite-range interlayer hopping amplitude. The conductance in turn scales as $\Delta^2$, Fig.~\ref{figD}d.  The solid line in Fig.~\ref{figD}d is $G_{el}\propto \exp(-L/l_{el})$ with the decay length $l_{el}\approx 0.25$ \AA. However, a few points show striking deviations, by up to a factor of 50.  This is because the electronic contribution is very sensitive to translation \cite{Mele1,McDonald1} as well as rotation \cite{EPAPS}.

\begin{figure}[hb]
\centering
\scalebox{1.0}[1.0]{
\includegraphics[width=3.5in]{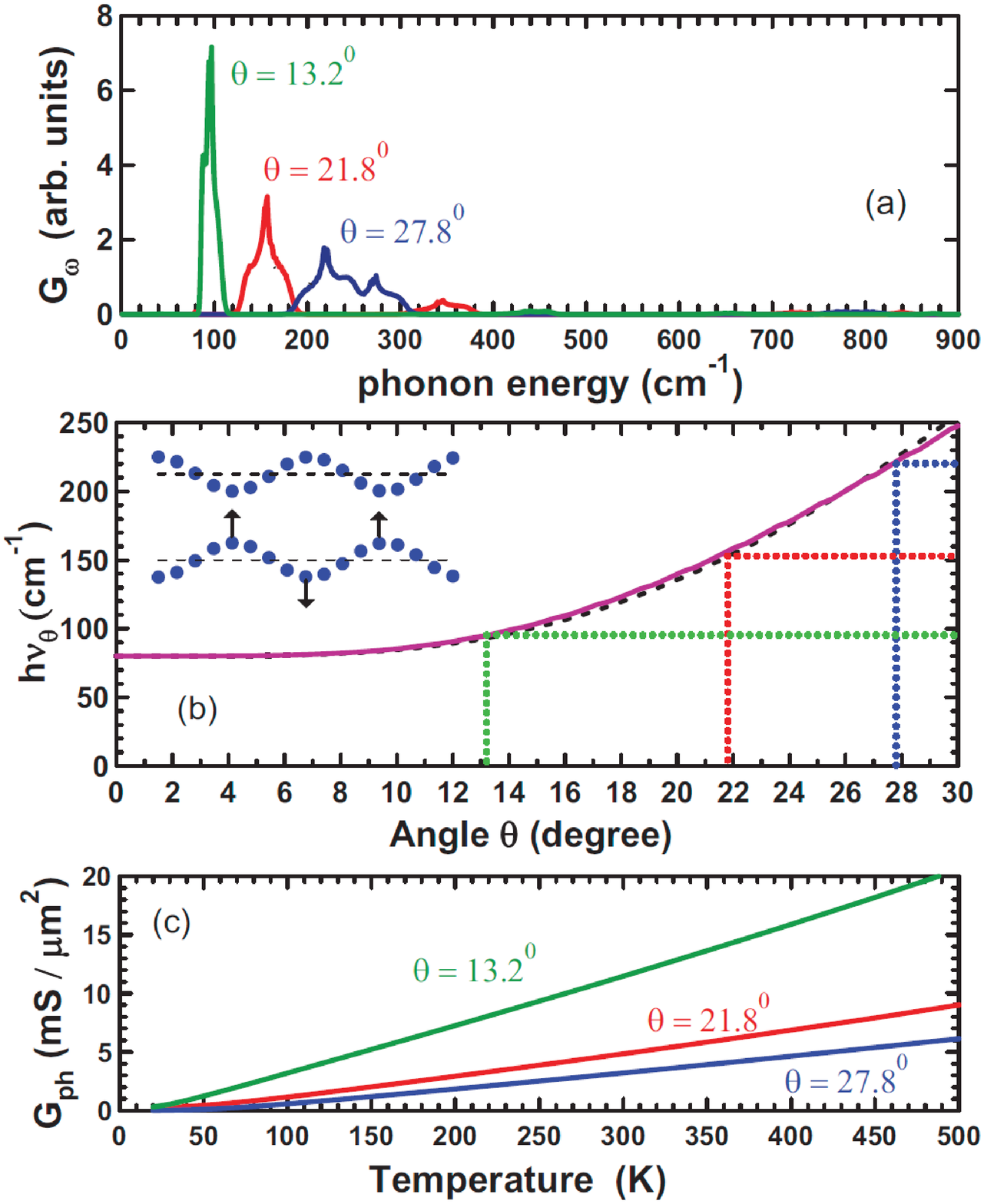}}
\caption{(Color  online)  (a) Relative contribution of phonons at different energies to the conductance. Peaks in $G_{\omega}$ clearly indicate the  dominant role of specific phonons.  (b) Energy of beating phonon mode at the wavevector $q_K$ connecting the rotated Dirac points  is shown as a function of rotation angle $\theta$. The dashed curve is a fit described in the text. As the angle increases, the relevant phonon energy increases.  The dashed lines show that for each angle, this phonon energy corresponds well to the structure in (a) for that angle.  Inset schematically illustrates beating mode geometry. (c) Temperature dependence of the low bias conductance in the same structures.}
\label{figE}
\end{figure}

In contrast to a single graphene layer, the bilayer has low-energy optical flexural phonons that can easily provide the momentum needed to couple the rotated $K$ points without Umklapp, so there is no exponential dependence on periodicity in Fig.~\ref{figD}(f) --- indeed, no obvious dependence on periodicity at all.   The full numerically results of Fig.~\ref{figD}(e) can be understood using a simplified model \cite{EPAPS} which considers coupling between the rotated Dirac points by the beating-mode phonon:
\begin{eqnarray}
G_{ph}\propto  g^2 E_F^2\frac{n_{\theta}}{\omega_{\theta}}
\label{eqscale}
\end{eqnarray}
where $\omega_{\theta}=\omega_{b}(q_K)$ is the  frequency of the beating mode $\omega_b$ at wavevector $q_K$ connecting the rotated Dirac points, $n_{\theta}$ is the thermal population of $\omega_{\theta}$ phonon, and the constant of proportionality is given in Ref.~\onlinecite{EPAPS}. This model has only one adjustable parameter, the effective electron-phonon matrix element $g$.
As expected, in Fig.~\ref{figD}(e) the simple model breaks down for small rotation angles, because $k_F$ can no longer be neglected relative to $q_K$.

Previous work \cite{Shallcross2,McDonald1} emphasized that the purely electronic conductance shows an extraordinary sensitivity to fine details of the model used in calculating the matrix elements.  This occurs because subtle changes in the analytic form of the interlayer interactions can radically change the asymptotic behavior at large $Q$, which controls Umklapp scattering. Once we include phonon scattering, this extraordinary sensitivity disappears, as does the strong sensitivity to lateral translation. (Details and further calculations are given in Ref.~\onlinecite{EPAPS}.)  Most importantly, phonons not only solve the ``numerical'' problems of hypersensitivity, but also solve the conceptual problems that arise in the small-scattering limit \cite{McDonald1}, so we can calculate absolute conductance without any phenomenological broadening.

The conductance depends sensitively on temperature, and also depends on bias voltage, as shown in Fig.~\ref{figG}a.  The behavior is particularly interesting at low temperature, Fig.~\ref{figG}b.  At sufficiently low temperatures, the phonon-mediated current shows a diode-like behavior, turning on at a sharply defined threshold voltage. The threshold voltage $V_{th}$ depends on the rotation angle. The sharpness of the current onset depends on the ratio $k_BT/eV_{th}$.

The general behavior at high biases can be understood from simple density of states argument (see Eq.~(\ref{eqscale})), which suggests scaling in the limit $E_F \gg k_BT$ \cite{Datta}: $G_{ph}\propto \left(eV \right)^{-1}  \int_{E_F-eV}^{E_F} E^2 dE=E_F^2-eVE_F+e^2V^2/3$. As shown in Fig.~\ref{figG}a this scaling works extremely well at high temperatures (dashed curves), reproducing the full numerical result with only one adjustable parameter chosen to match the zero-bias conductance  at each temperature.

The current is suppressed at low temperatures and low biases, because the required phonons cannot be excited either thermally or electronically.  As the bias becomes large enough for the carriers to excite a phonon, conductance starts to increase as shown in Fig.~\ref{figG}a. (A somewhat related effect has been suggested in tunneling from a metal tip to graphene \cite{ZhangSTM,WehlingSTM}.) The relevant phonon energy $\omega_{\theta}$ increases with the rotation angle, Fig.~\ref{figE}b, so a larger threshold bias is needed to allow phonon assisted tunneling as seen in Fig.~\ref{figG}b. As a result, the low-conductance plateau gets wider as $V_{th} \propto \omega_{\theta}$ and the ``on/off'' conductance ratio increases as $\propto 1/n_{\theta}$.

To see clearly which phonons contribute most to the total conductance, in Fig.~\ref{figE}a we plot the fraction of conductance arising from phonons at a specific energy.  In all three cases, the energy of the main peak corresponds to the beating mode connecting the rotated Fermi circles.  This is shown by the dashed lines in Fig.~\ref{figE}b, those phonon energies at 95 cm$^{-1}$, 150 cm$^{-1}$, 220 cm$^{-1}$ roughly coincide with the onsets of the phonon contribution to the conductance for the three structures calculated in Fig.~\ref{figE}a.  (The in-phase flexural phonon, whose energy vanishes at the $\Gamma$-point, does not contribute to the conductance.)

The dashed line in Fig.~\ref{figE}b shows a fit to the beating phonon mode phenomenological dispersion,
$\omega_{b}(q_K)= \left[ \omega_{\Gamma}^2+q_K^4 \kappa / \rho  \right]^{1/2}$, where
$\omega_{\Gamma}$ is zone center beating phonon frequency, $\rho$ is the graphene mass density, and  the single fitting parameter is the effective bending stiffness $\kappa$ \cite{ftnote1}.   The energy $\omega_{b}(q_K)$ of the phonon coupling Dirac points also sets the temperature scale for the crossover from a linear to a sublinear behavior, as shown in Fig.~\ref{figE}c.

At low rotation angles, the phonon-mediated conductance is expected to be larger than that in the calculations above due to the electronic coupling to the interlayer sliding mode \cite{Ferrari1}, and to any Fermi velocity renormalization \cite{Santos}.
We estimate \cite{EPAPS} that the current is underestimated by less than 10\%  for angles larger than $\theta>10^{\circ}$.

In conclusion, we have shown that phonon scattering is a dominant factor in the interlayer conduction of twisted  bilayers of graphene.  It relaxes the momentum conservation, eliminating the extraordinary sensitivity to rotation and translation of the layers, and it drastically changes the functional dependence.  Even at low temperature, phonons can dominate the conductance, provided that the bias across the layers is large enough to allow the excitation of a phonon during the tunneling. This effect leads to intriguing diode-like electrical characteristics with high ``on/off'' conductance ratio,  provided that other scattering mechanisms are small. These phenomena should be relevant in other two dimensional interfaces where current is suppressed by lateral momentum mismatch of the electronic states, which may be expected in novel tunneling devices \cite{Novoselov12}.

\bibliography{model}

\begin{thebibliography}{46}
\bibitem [{EPA()}]{EPAPS}%
  \BibitemOpen
  \href@noop {} {\bibinfo  {journal} {See pages 6-8 for additional technical
  details on our calculations and results}\ }\BibitemShut {NoStop}%
\bibitem [{ftn({\natexlab{a}})}]{ftnote2}%
  \BibitemOpen
  \href@noop {} {\bibfield  {journal} {\bibinfo  {journal} {The doping levels
  in both layers throughout the paper are assumed to be the same in the absence
  of bias voltage, while a bias shifts the Fermi energy $E_F$ of one layer by
  $eV$}\ } ({\natexlab{a}})}\BibitemShut {NoStop}%
\bibitem [{LJp()}]{LJparam}%
  \BibitemOpen
  \href@noop {} {\bibinfo  {journal} {The LJ parameters are determined by
  setting the van der Waals adhesion energy to $E_{a}=40$ meV per atom
  \cite{Lad} and interlayer spacing to 3.35 \AA}\ }\BibitemShut {NoStop}%
\bibitem [{Gel()}]{Geldetail}%
  \BibitemOpen
\bibfield  {journal} {  }\href@noop {} {\bibinfo  {journal} {We calculate
  $G_{el}$ from the square of the off-diagonal matrix elements
  \protect{\cite{Mele1,McDonald1}} connecting the Fermi circles in both
  layers}\ }\BibitemShut {NoStop}%
\end{thebibliography}

\newpage

\section{\label{append1} Analytical expression for $G_{ph}$}

The full numerically results of Fig.~1e of the main text can be understood using a simplified model.
The phonon must provide the momentum connecting the respective Fermi circles, $q_K \approx 2K\sin(\theta/2)$. Here $K=4\pi/(3a) \approx 17$ nm$^{-1}$, and we neglect the Fermi circle radius $k_F$ relative to the separation $q_K$.  The dominant contribution comes from the beating mode (i.e. the out-of-phase bilayer flexural mode).  Approximating the associated electron-phonon matrix element as momentum-independent, $\vert \langle\Psi_{k}^{s}|H_{e-ph}^{\mu}|\Psi_{k^{\prime}}^{s^{\prime}}\rangle \vert^2 \approx g^2\hbar/(2M_C\omega_{\theta})$, where $\omega_{\theta}=\omega_{b}(q_K)$ is the  frequency of the beating mode $\omega_b$ at wavevector $q_K$. Then the low bias conductance can be evaluated from  Eqs.~(1) and (2) of the main text:
\begin{align}
G_{ph}=2\pi e^2 g^2 n_{\theta} E_F^2A_0/(M_C\omega_{\theta}\pi^2\hbar^4v_F^4)\tag{S1}
\label{eqscale}
\end{align}
where $M_C$ is the carbon mass, $A_0=a^2\sqrt{3}/2$ is the area of the two atom cell, $v_F=\sqrt{3}ta/(2\hbar)\approx 10^8$ cm/s is the Fermi velocity, $n_{\theta}$ is the thermal population of  $\omega_{\theta}$ phonon, and $g$ is an effective (momentum averaged) electron-phonon constant.  With a single adjustable parameter $g\approx 0.34$ eV/\AA, Eq.~(\ref{eqscale}) accounts very well for the numerical result over a wide range of rotational angles as shown by the solid curve in Fig.~1(e) of the main text. As expected, the simple model breaks down for small rotation angles, because $k_F$ can no longer be neglected relative to $q_K$.

\section{\label{append2} Additional contributions to $G_{ph}$ at low rotation angles}

\begin{figure}[hb]
\includegraphics[width=3.5in]{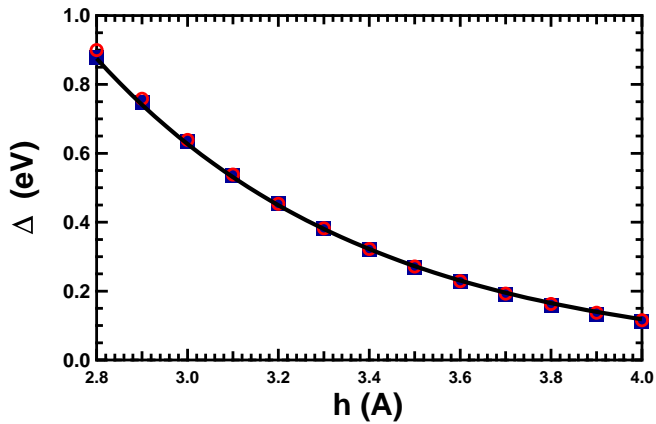}
\caption{(Color online) LDA calculations of the band splitting $\Delta$ in AB stacked bilayer graphene. $\Delta$ versus the interlayer distance $h$: from LDA calculations (blue squares), from the model Hamiltonian Eq.~(3) in the main text (red hollow circles), and an exponential fit: $\Delta=\Delta_0 \exp{((h_0-h)/\lambda)}$ with $\lambda=0.6$ \AA\ and  $\Delta_0=0.35$ eV at $h_0=3.35$ \AA.}
\label{figS1}
\end{figure}

At low rotation angles, the phonon-mediated conductance is expected to be larger than that in the calculations above due to the electronic coupling to the interlayer sliding mode, and to any Fermi velocity renormalization\cite{Santos}. First, we estimate the sliding mode contribution by assuming the electron-phonon coupling strength for the sliding mode and beating phonons are similar  $\frac{\partial t_{ij}}{\partial \xi_{ij}}\sim \frac{\partial t_{ij}}{\partial z_{ij}} $. Using acoustic sound velocity of $v_s=20 \times 10^3$ m/s and the zone center sliding mode energy in the bilayer \cite{Ferrari1}  $\hbar\omega_{s0}=31$ cm$^{-1}$ we can estimate the relevant phonon frequency as $\omega_{s}=\sqrt{\omega_{s0}^2+v^2_sq^2_K}$. According to Eq.~(\ref{eqscale}), the total conductance would scale as $G_{ph}(1+ \omega_{\theta}^2/\omega^2_{s})$.  This increases the conductance by only 7\% for $\theta=10^{\circ}$, increasing to 25\% and 65\% for $\theta=5^{\circ}$ and $3^{\circ}$ respectively. At the doping level used in Fig.~1 of the main text, the respective Fermi circles start to overlap at $\theta<3^{\circ}$.
To estimate the Fermi velocity renormalization $\tilde{v}_F$, we use Ref.~\cite{Santos}: $\tilde{v}_F/v_F=1-9[\tilde{t}_{\perp}/(\hbar v_Fq_K)]^2$, where $\tilde{t}_{\perp}=0.11$ eV.  The Fermi energy $E_F=\hbar \tilde{v}_F\sqrt{\pi n_s}$ used in Eq.~(\ref{eqscale}) will also be reduced for a fixed concentration $n_s=5 \times 10^{12}$ cm$^{-2}$, increasing the conductances by roughly a factor of $v_F^2/\tilde{v}_F^2\approx2.14$, $1.26$, and $1.06$ for $\theta=3^{\circ}$ , $5^{\circ}$, and $10^{\circ}$ respectively. Therefore, our results in Fig.~1e and 1f of the main text are expected to be quantitatively robust (within 10\%) for angles larger than $\theta>10^{\circ}$.  At smaller angles the phonon-mediated conductance is larger than calculated here, further enhancing the interlayer transport.

\begin{figure}[hb]
\includegraphics[width=3.5in]{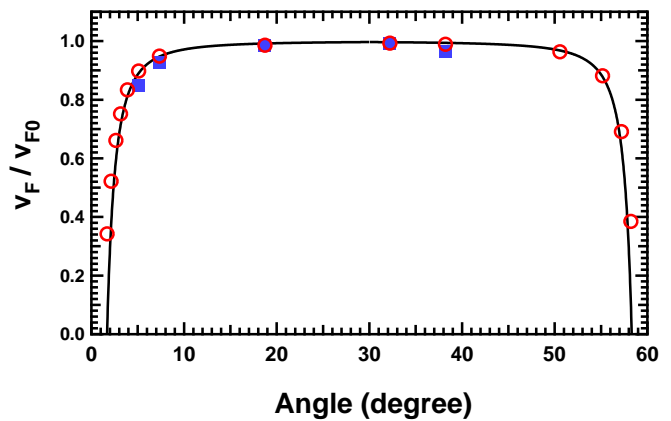}
\caption{(Color online) Tight-binding model test against LDA results for the velocity renormalization. The blue squares are results from Ref.~\protect{\cite{Trambly}} in structures from left to right: $(n, m)=(6, 7),$ (4, 5), (5, 9), (1, 3), and (1, 4). The hollow red circles are our tight-binding results in structures from left to right: (19, 20), (15, 16), (12, 13), (10,11), (8, 9), (6, 7), (4, 5), (5, 9), (1, 3),  (1, 4), (1, 10), (1, 20), (1, 35), (1, 55).   The black solid curve gives a perturbation theory result from Ref.~\protect{\cite{Santos}}: $\tilde{v}_F/v_F=1-9[\tilde{t}_{\perp}/(\hbar v_Fq_K)]^2$, where $\tilde{t}_{\perp}=0.11$ eV.}
\label{figS12}
\end{figure}

\section{\label{append3} Fitting of our tight-binding model to LDA results}

To obtain reasonable quantitative accuracy, we desire a model that correctly describes the dependence on both rotation and interlayer spacing, insofar as these are known from LDA calculations. We require that the model reproduce the LDA results for the dependence of $\Delta$ on layer spacing, which we calculated in the Bernal stacked bilayer as shown in Fig.~\ref{figS1}. Our model in Eq.~(3) of the main text contains four parameters: $t_{\perp}$, $\lambda_z$, $\lambda_{xy}$ and $\alpha$. We have used only three LDA values to fit them: $\Delta$ in Bernal stacked bilayer, interlayer spacing dependence in Bernal stacked bilayer (i.e. derivative $d\Delta/dh$ at $h=h_0$), and $\Delta$ value in $7 \times 7$ twisted bilayer. The model could be further improved by fitting to more LDA calculations. Nevertheless, we tested that our model describes reasonably well other LDA calculations, such as Fermi velocity renormalization shown in Fig.~\ref{figS12},  which were not used in fitting the parameters of the tight-binding model.

\begin{figure}[hb]
\includegraphics[width=3.5in]{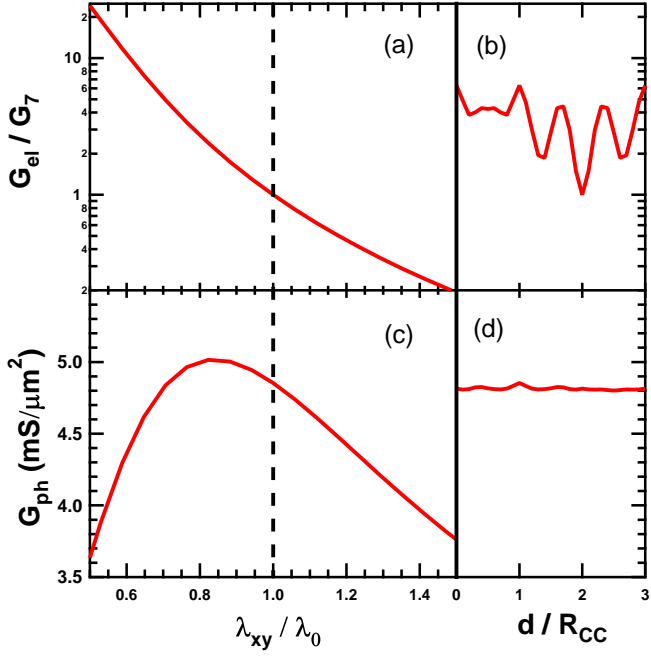}
\caption{(Color online) Sensitivity of electronic conductance $G_{el}$ and phonon mediated conductance $G_{ph}$ to the lateral cut-off distance $\lambda_{xy}$ and on the sliding shift $d$. Here $\lambda_0=1.7$ \AA \ \ and the sliding shift $d$ (in units of carbon bondlength $R_{CC}$) is along the C-C bond direction in the AA stacked bilayer after rotation, with $\theta= 21.8^{\circ}$. (a) and (b) show $G_{el}$ at $T=0$ K, while (c) and (d) show  $G_{ph}$ at room temperature and $E_F=0.26$ eV. Note, that $d=0$ corresponds to the structure where AA stacked bilayer is rotated around AA axes and translational shift $d=3R_{CC}$ gives an equivalent structure. Conventionally, twisted structures are obtained from the AB stacked bilayer rotated around the AB axes by a given angle, which would correspond to structure with  $d=2R_{CC}$ used for calculations in the main text. Structure with $d=R_{CC}$ corresponds to the conventional structure rotated by a complimentary angle $60^{\circ}-\theta$, i.e. $\theta= 38.2^{\circ}$  }
\label{figS3}
\end{figure}

\begin{figure}[hb]
\includegraphics[width=3.5in]{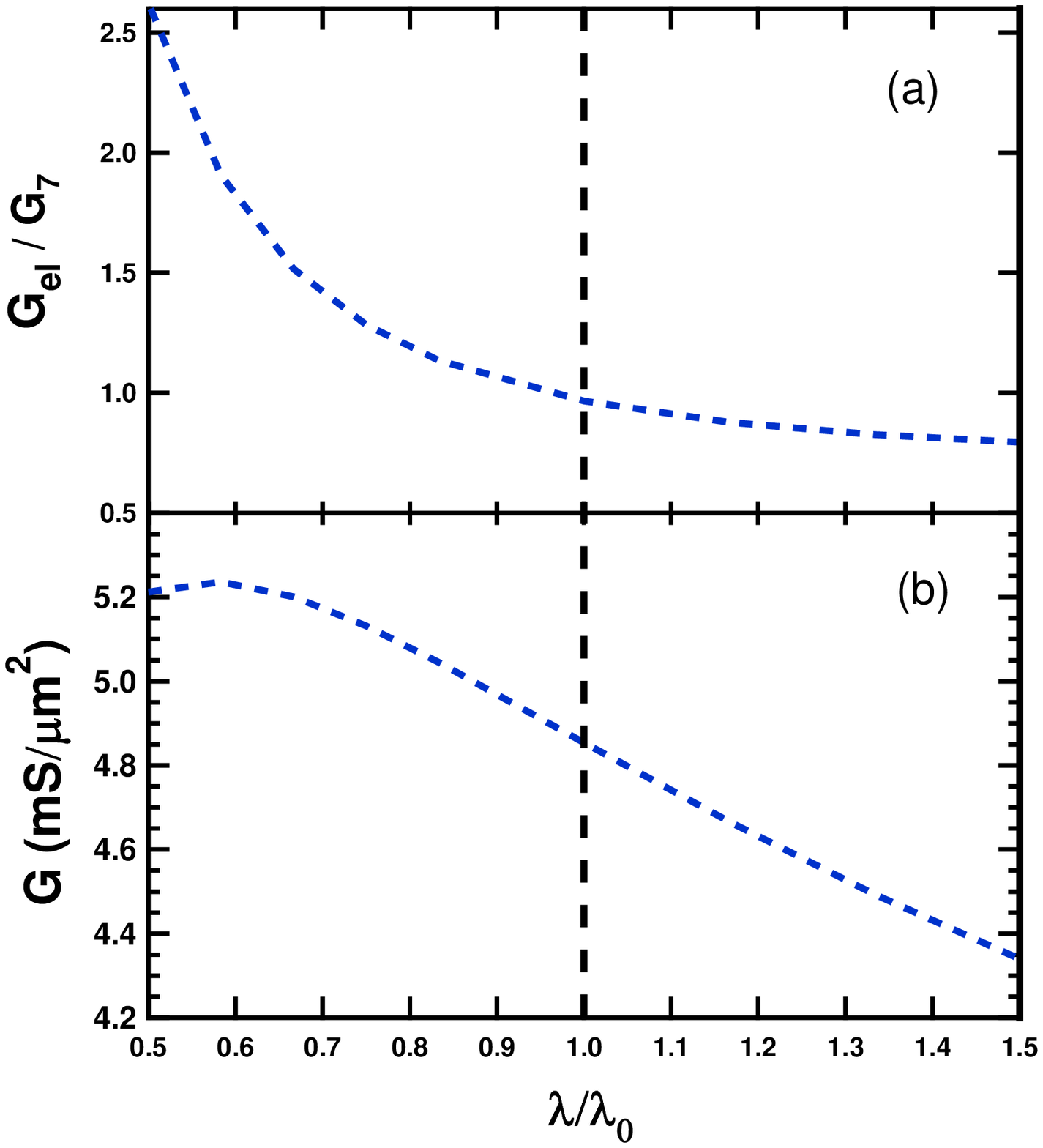}
\caption{(Color online) Sensitivity of electronic conductance $G_{el}$ and phonon mediated conductance $G_{ph}$ to the value of $\lambda$.
Here $\lambda_0=0.6$ \AA, see Eq.~(3) in the main text. \ \  (a) $G_{el}$ at $T=0$ K. (b) $G_{ph}$ at room temperature and $E_F=0.26$ eV. }
\label{figS4}
\end{figure}

We emphasize that the qualitative points of our paper could have been obtained with the previous models.  We introduce a new model simply as a tool to get reasonable accuracy for the electron-phonon coupling, since the previous models were not designed to give the correct interlayer distance dependence.  In Figs.~\ref{figS3} and \ref{figS4} we show that variation of the parameters of the model by as much as a factor of two does not change the phonon mediated conductance by more than 30\%. This can be contrasted with the electronic conductance, which is far much more sensitive to the parameters of the tight-binding model.  The same is true of the sensitivity to lateral sliding of one layer with respect to the other, also shown in Figs.~\ref{figS3} and \ref{figS4}.

\begin{figure}[hb]
\includegraphics[width=3.5in]{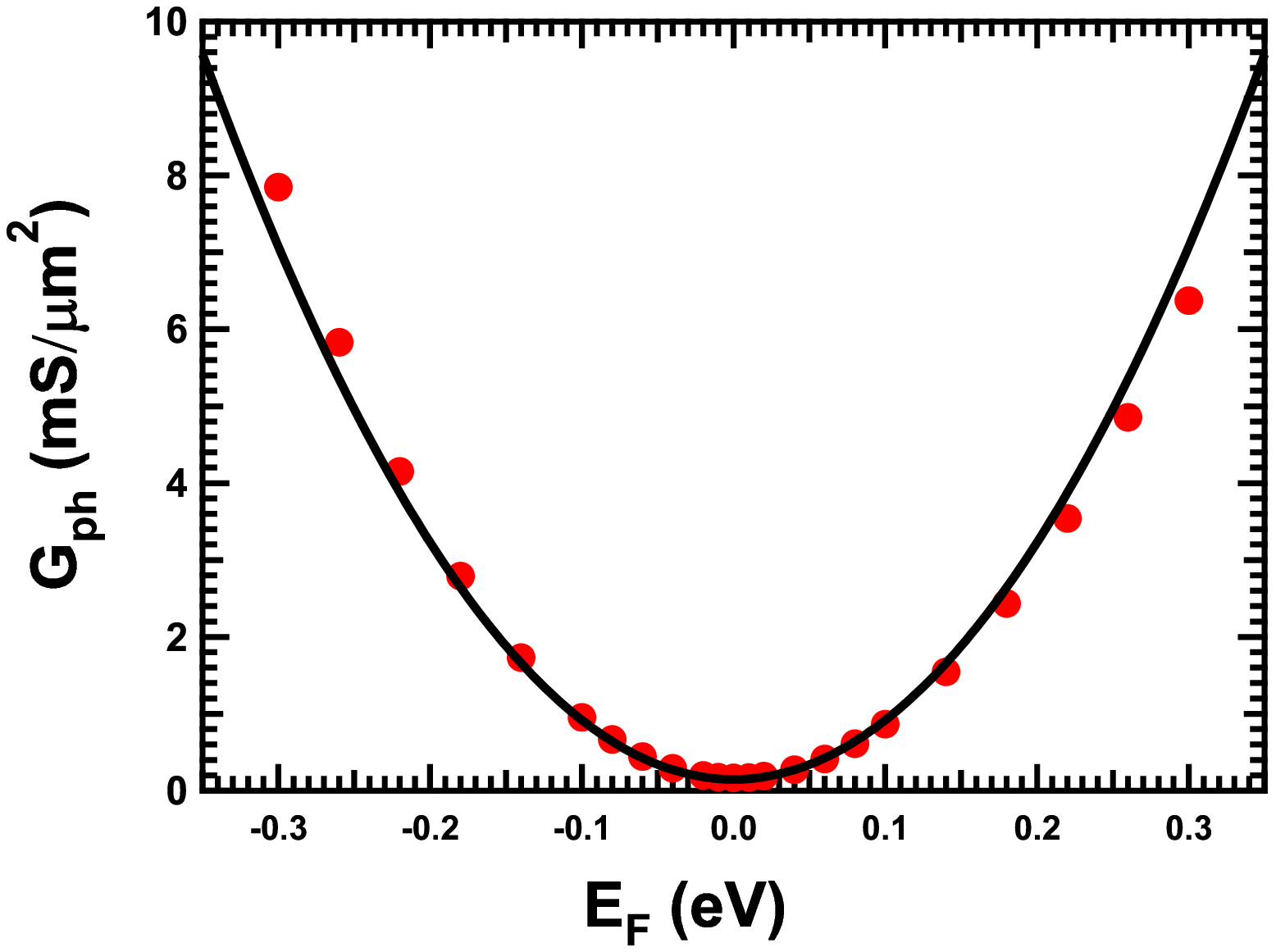}
\caption{(Color online) Fermi energy dependence of the phonon mediated conductance. $G_{ph}$  in $\theta \approx 21.8^{\circ}$ bilayer structure at T=300 K (red circles). The black solid curve is a parabolic fit motivated by the approximation of momentum independent electron-phonon matrix elements, i.e. Eq.~(\ref{eqscale}). The single parameter fit does not capture the slight electron-hole asymmetry which comes from the long-range nature of the interaction in Eq.~(3) of the main text. The small residual conductance at $E_F=0$ is due to the thermal electron-hole pair excitations.}
\label{figS2}
\end{figure}

Finally, in Fig.~\ref{figS2} shows that the long range nature of the tight-binding model in Eq.~(3) of the main text introduces relatively small electron-hole asymmetry, so all the results in the main text are reported for the electron branch.



\end{document}